\documentclass{aastex62}

\usepackage{color}
\usepackage{natbib}
\usepackage{diagbox}

\newcommand{\teff}{$T_{\rm eff}$}
\newcommand{\logg}{$\log g$}
\newcommand{\feh}{[Fe/H]}
\newcommand{\kepler}{\emph{Kepler}}

\usepackage{nicefrac}
\usepackage{multirow}
\DeclareUnicodeCharacter{02BC}{'}
\shorttitle{giant li evolution}
\shortauthors{Jinghua Zhang et al.}

\begin{document}

\title{Lithium Evolution of Giant Stars Observed by LAMOST and \kepler{}}

\author[0000-0002-2510-6931]{Jinghua Zhang}
\affiliation{CAS Key Laboratory of Optical Astronomy, National Astronomical Observatories, Chinese Academy of Sciences, Beijing 100101, People's Republic of China}

\author[0000-0002-0349-7839]{Jian-Rong Shi}
\email{sjr@nao.cas.cn}
\affiliation{CAS Key Laboratory of Optical Astronomy, National Astronomical Observatories, Chinese Academy of Sciences, Beijing 100101, People's Republic of China}
\affiliation{School of Astronomy and Space Science, University of Chinese Academy of Sciences,  Beijing 100101, People's Republic of China}

\author[0000-0002-8609-3599]{Hong-Liang Yan}
\affiliation{CAS Key Laboratory of Optical Astronomy, National Astronomical Observatories, Chinese Academy of Sciences, Beijing 100101, People's Republic of China}
\affiliation{School of Astronomy and Space Science, University of Chinese Academy of Sciences,  Beijing 100101, People's Republic of China}
\email{hlyan@nao.cas.cn}

\author[0000-0003-3020-4437]{Yaguang Li}
\affiliation{Sydney Institute for Astronomy (SIfA), School of Physics, University of Sydney, NSW 2006, Australia}
\affiliation{Stellar Astrophysics Centre, Department of Physics and Astronomy, Aarhus University, Ny Munkegade 120, DK-8000 Aarhus C, Denmark}

\author[0000-0003-4972-0677]{Qi Gao}
\affiliation{CAS Key Laboratory of Optical Astronomy, National Astronomical Observatories, Chinese Academy of Sciences, Beijing 100101, People's Republic of China}
\affiliation{School of Astronomy and Space Science, University of Chinese Academy of Sciences,  Beijing 100101, People's Republic of China}

\author[0000-0002-6647-3957]{Chun-Qian Li}
\affiliation{CAS Key Laboratory of Optical Astronomy, National Astronomical Observatories, Chinese Academy of Sciences, Beijing 100101, People's Republic of China}
\affiliation{School of Astronomy and Space Science, University of Chinese Academy of Sciences,  Beijing 100101, People's Republic of China}

\author[0000-0002-3672-2166]{Xianfei Zhang}
\affiliation{Department of Astronomy, Beijing Normal University, Beijing 100875, People's Republic of China}

\author{Shuai Liu}
\affiliation{CAS Key Laboratory of Optical Astronomy, National Astronomical Observatories, Chinese Academy of Sciences, Beijing 100101, People's Republic of China}
\affiliation{School of Astronomy and Space Science, University of Chinese Academy of Sciences,  Beijing 100101, People's Republic of China}

\author{Shaolan Bi}
\affiliation{Department of Astronomy, Beijing Normal University, Beijing 100875, People's Republic of China}

\author[0000-0002-8980-945X]{Gang Zhao}
\affiliation{CAS Key Laboratory of Optical Astronomy, National Astronomical Observatories, Chinese Academy of Sciences, Beijing 100101, People's Republic of China}
\affiliation{School of Astronomy and Space Science, University of Chinese Academy of Sciences,  Beijing 100101, People's Republic of China}

\author[0000-0002-1424-3164]{Yan Li}
\affiliation{School of Astronomy and Space Science, University of Chinese Academy of Sciences,  Beijing 100101, People's Republic of China}
\affiliation{Yunnan Observatories, Chinese Academy of Sciences, Kunming 650216, Peopleʼs Republic of China}
\affiliation{Key Laboratory for Structure and Evolution of Celestial Objects, Chinese Academy of Sciences, Kunming 650216, Peopleʼs Republic of China}
\affiliation{Center for Astronomical Mega-Science, Chinese Academy of Sciences, Beijing, 100012, Peopleʼs Republic of China}




\begin{abstract}
Mapping lithium evolution for evolved stars will provide restrictions and constraints on the fundamental stellar interior physical processes, which further shed light on our understanding of the theory of stellar structure and evolution. Based on a sample of 1,848 giants with known evolutionary phases and lithium abundances from the LAMOST-\kepler{} and LAMOST-\emph{K}2 fields, we construct mass-radius diagrams to characterize the evolutionary features of lithium. The stars at red giant branch (RGB) phase show natural depletion along with their stellar evolution, particularly, there is no obvious crowd stars with anomalously high Li abundances near the bump. Most of the low-mass stars reaching their zero-age sequence of core-helium-burning (ZAHeB) have Li abundances around $\sim1.0$\,dex, which show an increase of Li abundance by $\sim0.6$\,dex compared to the stars above the bump of RGB. This suggests the helium flash can be responsible for moderate Li production. While for super Li-rich stars, some special mechanisms should be considered during helium flash. Other scenarios, such as merger, could also be interpretations given the Li-rich stars can be found at anytime during the steady state phase of core He-burning. During the core He-burning (HeB) phase, there is no indication of obvious lithium depletion. 

\end{abstract}


\section{Introduction} \label{sec:intro}

The surface lithium abundances display various patterns for stars of different types as well as at different evolutionary stages.  
Lithium is easily destroyed by nuclear reactions in stars, and it is sensitive to mixing events connecting the envelope with warm layers where it easily undergoes p-captures. Thus, the surface Li has experienced depletion over time for stars with convective envelope \citep{Twarog2020MmSAI,Randich2021}. Therefore, the signatures of Li are good manifestations of subtle physical processes beneath the surface, and provide key information on inside stellar structure and evolution. 

During the past few decades, the increasing observational results have confirmed the theoretical predictions on the depletions of Li in giant stars \citep[e.g.][]{Lind2009,Deliyannis2019,Twarog2020MmSAI}. However, due to the difficulties of classifying evolutionary stages, especially separating core-helium-burning (HeB) from red giant branch (RGB)  bump stars by traditional approach, the evolutionary features of Li from RGB to HeB stage are still in the dark. Moreover, the nature of Li-rich giants found in a small number of objects is still beyond the standard stellar evolution theory \citep[e.g.][]{Wallerstein1982,Yan2018,Smiljanic2018,Gao2019ApJS,Casey2019ApJ,Deepak2019,Martell2021}. The physical mechanisms responsible for those Li-rich giants are under debate \citep[e.g.][]{Charbonnel2010,Kumar2020NatAs,Schwab2020ApJ,Zhang2020ApJ,Mori2021MN,Yan2021NatAs,Singh2021arXiv}, which further complicates Li evolution in evolved stars. 

Fortunately, the asteroseismic analysis for a mass of giants in the  \kepler{} \citep{Borucki2010} and K2 \citep{Howell2014PASP} fields has uncovered their evolutionary stages \citep[e.g.][]{Stello2013ApJ,Mosser2015,Vrard2016,Hon2018MN,Zinn2020ApJS}. Particularly, based on the asteroseismic scaling relations, the positions of RGB bump as well as the zero-age sequence of HeB (ZAHeB) stars could be well defined in mass-radius diagram \citep{Li2021MNRAS}. \citet{Deepak2021} explored evolution of lithium in giant stars based on $\sim$200 seismic data. Based on the distribution of $A_{\rm Li}$ as a function of mean large frequency separation ($\Delta\nu$), they found no indication of Li enrichment near the luminosity bump. \citet{Martell2021} explored the properties of 1262 Li-rich giant stars from the GALAH \citep{DeSilva2015} and K2 surveys, and found that the less massive primary red clump stars are 1.5 times as likely to be super lithium-rich as the more massive secondary red clump stars. Differing from the result shown in \citet{Deepak2021}, \citet{Martell2021} showed that there is a concentration of lithium-rich stars near the luminosity of the red giant branch bump. Recently, there are thousands of stars in the \kepler{} and K2 fields have been observed by the Large Sky Area Multi-Object Fibre Spectroscopic Telescope (LAMOST) spectroscopic survey \citep{Zhao2006, Cui2012, Zhao2012}, and their surface Li abundances have been presented recently \citep{Gao2019ApJS,Gao2021ApJS}. Combining these data sets, it is possible for us to investigate the signatures of Li abundances for stars evolved from RGB to HeB phase. Also, it would shed light on the Li production in evolved stars, and give more detailed insight into the nature of the stellar evolution theory. 

In this letter, we characterized signatures of Li abundances for giant stars at different evolutionary stages which were observed by both LAMOST and \kepler{} (K2 included). In Section~ \ref{sec:data}, we introduce the sample selection. In Section~\ref{sec:Li_HR}, we estimate the mass and radius for our sample stars, and characterize the evolutionary features of Li abundances in different stages. We summarise our results in Section~\ref{sec:conclusion}.

\section{Stellar samples} \label{sec:data}

\citet{Gao2019ApJS} searched for Li-rich giants using the LAMOST low-resolution spectra, and they found 10,535 giant stars with $A_{\rm Li}\geq1.5$\,dex. This catalog provided the updated largest sample of Li-rich giants. Very recently, \citet{Gao2021ApJS} further determined Li abundances for a sample of 165,479 stars based on the LAMOST medium-resolution spectra. We thus considered the two catalogs as our initial targets, and cross-matched the combined catalog with these of \citet{Yu2018ApJS} and \citet{Zinn2020ApJS}. The latter two catalogs provided the frequency of maximum power ($\nu_{\rm max}$), the mean large frequency separation ($\Delta\nu$) and the classification of evolutionary stages for samples of 16,094 giants in the \kepler{} field  and of 4,395 giants in the K2 fields, respectively. A total of 1,848 giant stars are finally selected, including 904 RGB stars and 944 HeB stars. Among them, there are 30 stars that have been detected Li abundances both in the low- and medium-resolution observational modes, and the Li results from medium-resolution modes are adopted in this work. Because of the way these stars were selected, we do not try to address the detection fractions of Li-rich stars here. However, we do note that Li was detected in $\sim$80\% of the sample, with no strong trends in detection fraction as a function of mass, metallicity, or evolutionary state likely to bias our analysis.

\section{Li abundances in \emph{M-R} diagram} \label{sec:Li_HR}

It has been proved that the asteroseismic parameters ($\Delta\nu$, $\nu_{\rm max}$) are tightly related with stellar fundamental parameters, and the stellar masses and radii can be estimated by the asteroseismic parameters with an addition of effective temperature (\teff), which are also known as the asteroseismic scaling relations \citep[see, e.g.][and references therein]{Stello2008,Kallinger2010}. These relations have also been confirmed having remarkable constraining powers with intrinsic scatter of mass and radius being limited to $\sim1.7$ and $\sim0.4$ per cent, respectively \citep{Li2021MNRAS}. Thereby, taking advantage of these asteroseismic scaling relations, we can construct a mass-radius diagram (hereafter, \emph{M-R} diagram) instead of traditional Hertzsprung–Russell diagram to investigate the evolutionary features of Li abundances in evolved stars.
We calculate the mass and radius of our sample following
  \begin{equation}\label{1}
    \frac{M}{M_{\odot}}\approx \bigg(\frac{\nu_{\rm max}}{f_{\nu_{\rm max}}\nu_{\rm max,\odot}}\bigg)^3\bigg(\frac{\Delta\nu}{f_{\Delta\nu}\Delta\nu_{\odot}}\bigg)^{-4}\bigg(\frac{T_{\rm eff}}{T_{\rm eff,\odot}}\bigg)^{3/2},
  \end{equation}
   \begin{equation}\label{2}
    \frac{R}{R_{\odot}}\approx \bigg(\frac{\nu_{\rm max}}{f_{\nu_{\rm max}}\nu_{\rm max,\odot}}\bigg)\bigg(\frac{\Delta\nu}{f_{\Delta\nu}\Delta\nu_{\odot}}\bigg)^{-2}\bigg(\frac{T_{\rm eff}}{T_{\rm eff,\odot}}\bigg)^{1/2},
  \end{equation}
where, $\nu_{\rm max,\odot} = 3090\,\mu$Hz, $\Delta\nu_{\odot} = 135.1\,\mu$Hz, and $T_{\rm eff,\odot} = 5777$\,K \citep{Huber2011ApJ}. The parameters $f_{\nu_{\rm max}}$ and $f_{\Delta\nu}$ are correction factors. Following \citet{Sharma2016}, the factor $f_{\Delta\nu}$ is a function of stellar atmospheric parameters (\teff, \logg, \feh) and evolutionary phase, and can be determined by interpolation in grids of models for $-3 <$ [Fe/H] $< 0.4$ and $0.8 < M/M_{\odot} < 4.0$. Since the factor $f_{\nu_{\rm max}}$ is difficult to be determined theoretically \citep{Belkacem2011}, it was set to be 1.0 in this work. The stellar atmospheric parameters were adopted from \citet{Gao2019ApJS} and \citet{Gao2021ApJS}. The $\Delta\nu$ and $\nu_{\rm max}$ for stars in the K2 fields were adopted from the mean values of different pipelines \citep{Zinn2020ApJS}, while for stars in the \kepler{} field they were determined via SYD pipeline \citep{Yu2018ApJS}. The small differences in mass and radius determined by different pipelines results would not affect us mapping lithium evolution.

\subsection{Li abundances in RGB stars} \label{sec:RGB}
\begin{figure}
\figurenum{1}
\epsscale{0.7}
\plotone{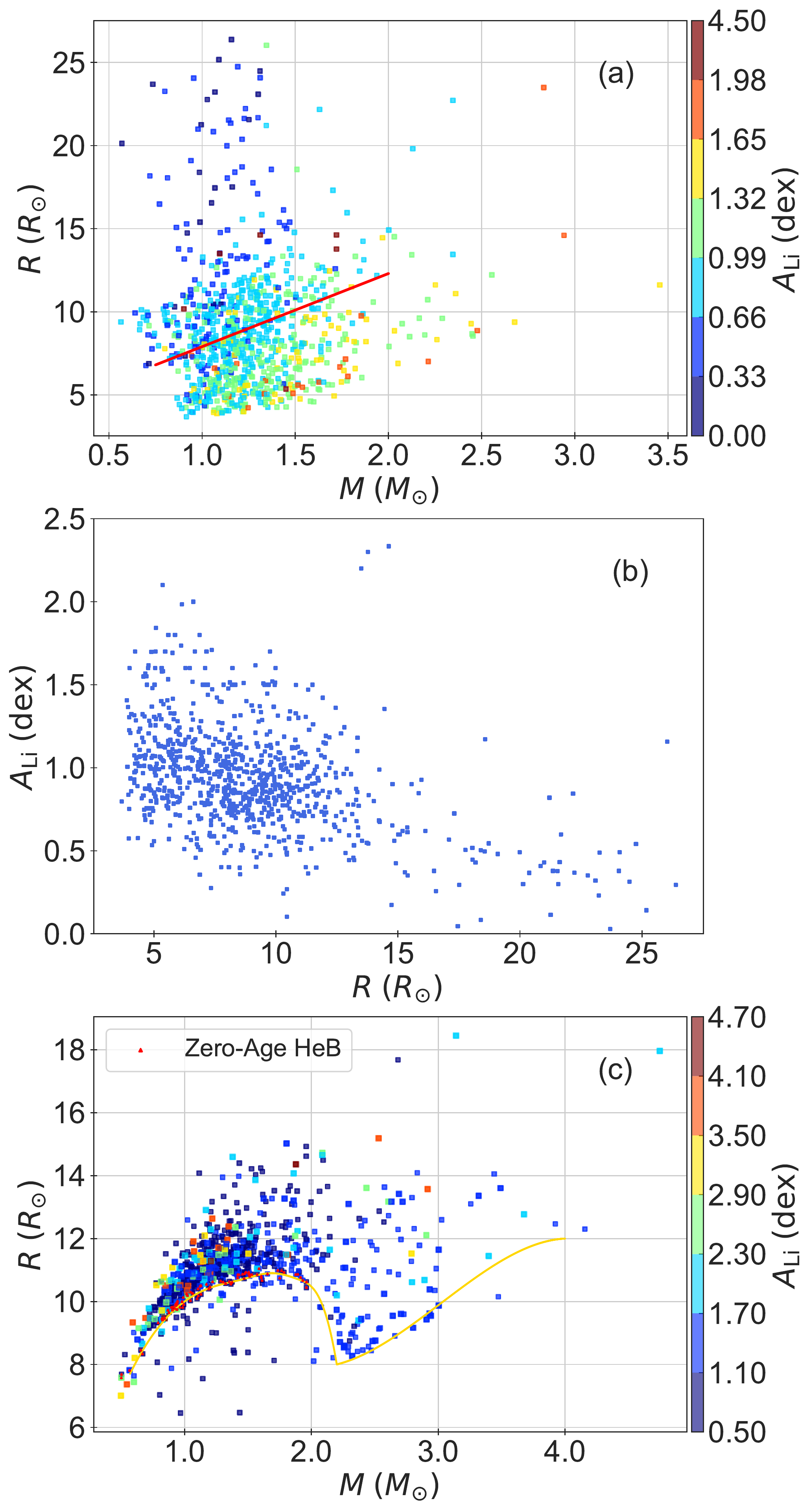}
\caption{Radius versus mass for RGB (upper) and HeB (lower) stars with color-coded by Li abundance, as well as radius versus Li abundance for low-mass RGB stars (middle). The red line in upper panel denotes the RGB bump. the gold spine in lower panel denotes the edge of ZAHeB stars and the red triangles denote the stars locating within reach of ZAHeB edge. \label{fig:Li_HR}}
\end{figure}

To better understand the evolutionary features of Li abundances and the origin of Li enhancement in evolved stars, we separate HeB from RGB for our sample stars, and characterize Li abundances for each population. Figure~\ref{fig:Li_HR} shows the \emph{M-R} diagrams of the RGB (panel a) and HeB (panel c) stars in our sample with color-coded by Li abundances, as well as the Li abundance versus radius for low-mass RGB stars (panel b). The red line in panel a defines the RGB bump, while the gold spline in panel c indicates the zero-age HeB (ZAHeB) edge \citep[see][for details]{Li2021MNRAS}. 

As shown in Figure~\ref{fig:Li_HR}b, for the majority of low-mass ($\emph{M} < 2\emph{M}_{\odot}$) RGB stars, the Li abundances decrease gradually by $\sim2.0$ dex when their radii expand by a factor of $\sim5$. This feature indicates the consequence of Li depletion along with stellar evolution, which has been predicted by the standard stellar evolution theory \citep[e.g.][]{Iben1967ApJ}. As shown in Figure~\ref{fig:Li_HR}a, there is no obvious crowd stars with anomalously high Li abundances near the bump. Considering the definition of $A_{\rm Li}\geq 1.5$\,dex for Li-rich stars, the total number of these stars among low-mass RGB accounts for 45, while only five of them evolve past the bump. This result implies that the RGB bump may be not the key to Li enhancement of RGB stars. The similar result was reported by \citet{Deepak2021}, who found no indication of Li enrichment near the luminosity bump. As suggested by previous studies \citep[e.g.][]{Yan2021NatAs}, these Li-rich stars may possess higher $A_{\rm Li}$ than their counterparts at main-sequence stage, they could still reserve relatively high $A_{\rm Li}$ after finishing the FDU process. Therefore, the Li-rich RGB stars might be a natural consequence of Li depletion along with stellar evolution. Given that the depletion of Li by FDU process is generally with a factor of $\sim60$ \citep{Iben1967ApJ}, if we take a low-mass sample star with the highest $A_{\rm Li}$ of $\sim2.4$\,dex below the bump, the $A_{\rm Li}$ would be up to $\sim4.2$\,dex at the main-sequence stage. Such stars have been reported by previous studies \citep[e.g.][]{{Li2018ApJ,Gao2021ApJS}}. The external scenario, on the other hand, such as engulfment of a substellar component \citep[e.g.][]{Alexander1967,Siess1999MN}, can not be excluded, which is in accordance with the observed Li abundances for our Li-rich RGB stars. A simulation on engulfment of a Jovian planet could result in a typical upper limit of $A_{\rm Li}\sim 2.2$\,dex \citep{Aguilera2016}, which is sufficient for Li production in most of Li-rich star below the bump in our sample.

As the mass increases, we found that Li abundances are comparably higher. Although the intermediate-mass RGB stars show Li depletion as that of the low-mass RGB objects, the depletion is moderate. In fact, the thinner convective envelope in intermediate-mass RGB stars usually accompany with relatively lower temperature in the deep convection zone and shorter time available for deep mixing that the destruction of Li is dependent on. As a result, high Li abundance can be easily found among the intermediate-mass RGB stars.

\subsection{Li abundances in HeB stars} \label{sec:HeB}
After leaving the tip of the RGB (tip-RGB) phase, stars evolve down in luminosity
and begin core helium burning (HeB), which are also known as the red clump phase. The low-luminosity edge defines the beginning of the red clump, which is also called the zero-age HeB (ZAHeB) phase. The edge is remarkable in H-R diagram as well as in \emph{M-R} diagram \citep{Girardi2016,Li2021MNRAS}, which makes it feasible to look into the evolutionary features of Li abundance among HeB stars. In Figure~\ref{fig:Li_HR}c, this edge is visible as the lower boundary.
\begin{figure}
\figurenum{2}
\epsscale{0.7}
\plotone{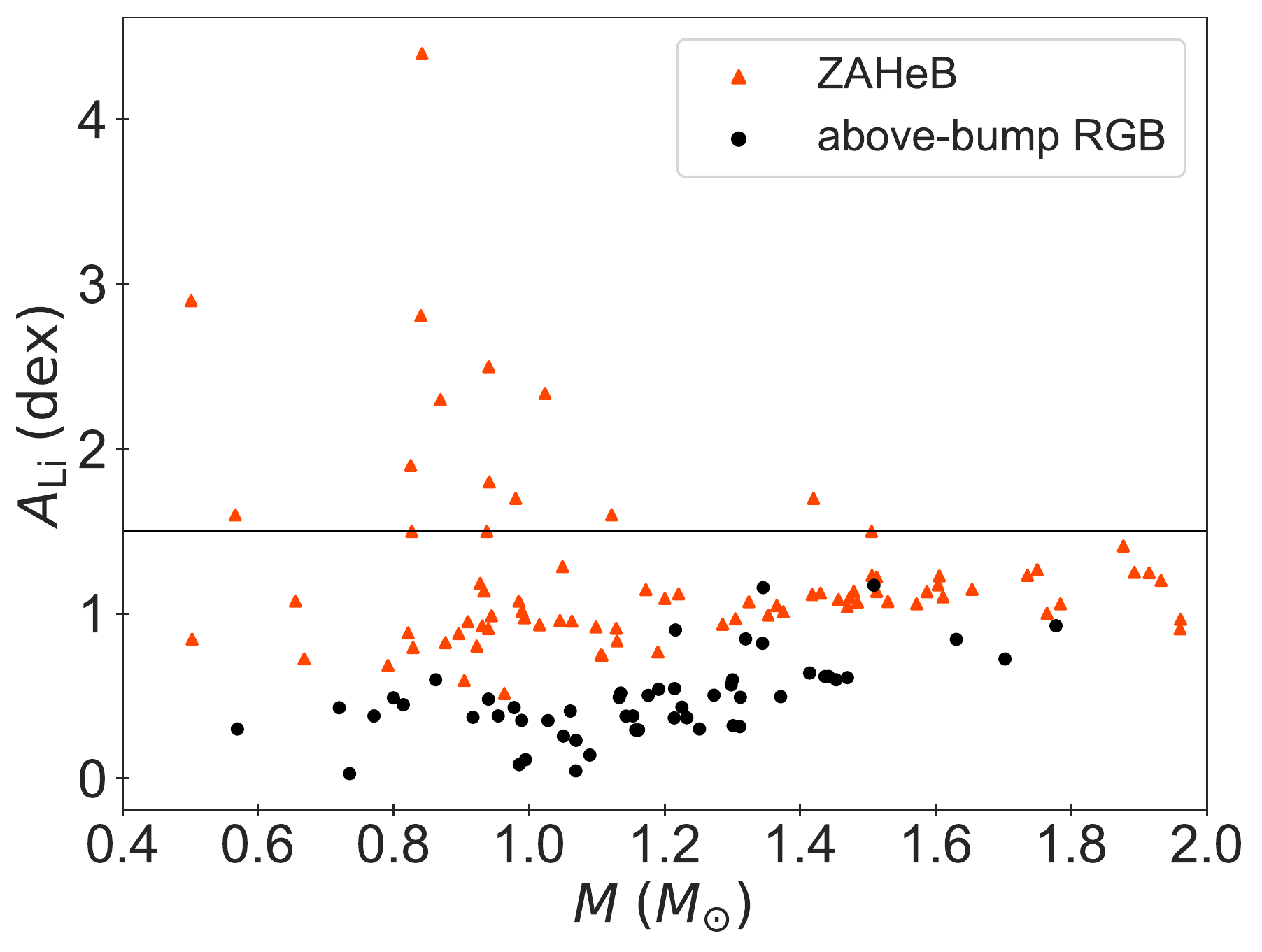}
\caption{Li abundances versus mass for stars locating above the RGB bump and HeB stars within reach of ZAHeB edge. The horizontal line denotes the lower limit of Li enhancement. \label{fig:Li_in_tipRGB_ZAHeB}}
\end{figure}

As shown in Figure~\ref{fig:Li_HR}c, the HeB stars possess high frequency of Li-rich stars whose Li abundances are higher than the primordial abundance \citep{Cyburt2016}. Moreover, most of these stars are dominated by low-mass stars. Interestingly, the radii of these stars are distributed over a wide range and do not show any significant feature,  which indicates that the super Li-rich stars can be found at anytime during the steady state phase of core-helium burning.

According to some recent studies, the helium-flash-induced mixing event occurs between tip-RGB and HeB phases, which is proposed to be responsible for Li enhancement among low-mass HeB stars \citep[see,][and references therein]{Schwab2020ApJ, Kumar2020NatAs}. Since the low-mass ZAHeB stars could be well defined for our sample, one would expect to see a significant increase of $A_{\rm Li}$ for these stars in comparison with that for stars at or near tip-RGB based on the scenario of helium-flash-induced mixing. Considering the asteroseismic measurements and inferences become less certain at large radii \citep{Pinsonneault2018}, we further select a subset of low-mass RGB stars with their radii greater than $15\emph{R}_{\odot}$ (Figure~\ref{fig:Li_HR}a), and we consider them as above-bump RGB stars. Then, we select the other subset of low-mass HeB stars locating within reach of ZAHeB edge (Figure~\ref{fig:Li_HR}c). We present the Li abundance versus mass for the two subsets in Figure~\ref{fig:Li_in_tipRGB_ZAHeB}. It can be seen that most ZAHeB stars have higher Li abundances than that of stars above the bump. The median $A_{\rm Li}$ increases by $\sim0.6$ dex from post-bump to ZAHeB phase. This increment is very close to the result given by the model prediction after including the helium-flash-induced mixing \citep[see their Fig 3 in][]{Schwab2020ApJ}, which means that the scenario of helium-flash-induced mixing indeed contributes Li production for most low-mass HeB stars. The increment shown in our sample provides an observational evidence for the Li production due to helium-flash-induced mixing. However, the $A_{\rm Li}$ of the majority of ZAHeB stars are close to 1.0\,dex, which hardly meet the traditional standard of Li-rich giants. For these Li-rich low-mass HeB stars, we suggest that some special physical mechanisms should be needed during the helium flash so that abnormal enhancement of Li could be triggered. Another possible interpretation for these Li-rich low-mass  HeB stars is the merger of a helium white dwarf (HeWD) with an RGB star \citep{Zhang2020ApJ}. The HeWD–RGB merger could result a Li-rich HeB star at anytime during the HeB stars' steady state phase of core He-burning \citep{Zhang2020ApJ}, which is compatible with the observational distribution of our sample. \citet{Yan2021NatAs} showed that the HeWD–RGB merger scenario generally agrees with the observed Li abundances in their sample stars. Our results do not support the proposal of \citet{Singh2021arXiv}, who used variations of asymptotic gravity-mode period spacing ($\Delta\Pi$) as tracer of duration time for core helium-burning and found that the super Li-rich giants are almost exclusively young HeB stars. Although in this work we could not define the upper edge of HeB in the \emph{M-R} diagram, the Li-rich low-mass HeB stars with large radii are supposed to be evolved HeB.

\begin{figure}
\figurenum{3}
\epsscale{0.8}
\plotone{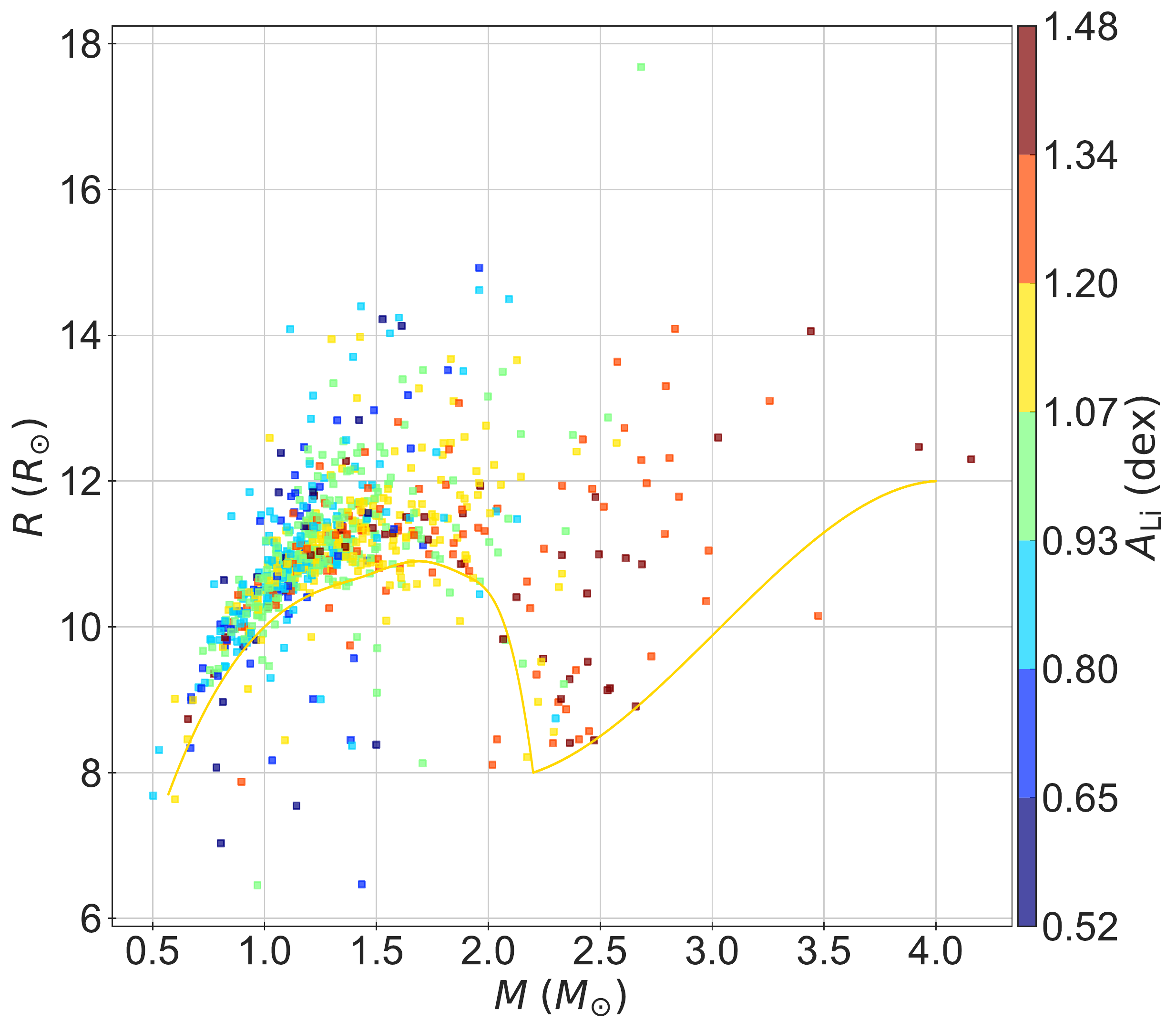}
\caption{Same as Figure~\ref{fig:Li_HR}c, but showing HeB stars with $A_{\rm Li} < 1.5$. \label{fig:normalLi_in_RC}}
\end{figure}

The features of Li-rich HeB stars indeed provide clues to understand stellar internal physical processes, however, they are just a small fraction of evolved stars. To give a more explicit picture for Li evolution, we here focused on those HeB stars with $A_{\rm Li}$ less than 1.5\,dex, which we referred as Li-normal HeB stars. We plot the \emph{M-R} diagram only for the Li-normal HeB stars in Figure~\ref{fig:normalLi_in_RC}. On one hand, it can be seen that the Li abundances strongly relate to the stellar masses, with higher mass stars possessing higher Li abundances. As a whole, the intermediate-mass stars have a median $A_{\rm Li}$ of 1.3\,dex, which is 0.3\,dex higher than that of the low-mass stars. This relation is also attributed to the thinner convective envelope in higher mass stars, which weakens the depletion of Li. On the other hand, there is no obvious sign of decreasing $A_{\rm Li}$ along with radii, this feature can be justified for that stars at red clump phase would not be affected by further mixing, because the retreat of convective envelope make it remaining rather far from the H-burning shell at this evolutionary stage \citep{Palmerini2011ApJ}. 

\section{conclusions} \label{sec:conclusion}
In this letter, we investigate the evolutionary features of Li for a sample of stars evolved from RGB to HeB stage in the LAMOST-\kepler{}/K2 project. The RGB stars show a Li depletion feature along with stellar evolution. The Li-rich RGB stars mainly locate below the bump, and there is no obvious crowd stars with anomalously high Li abundances near the bump, which indicates that Li-rich RGB stars might be a natural consequence of Li depletion along with stellar evolution. We found an immediate increase of Li abundance by $\sim0.6$\,dex for stars between post-bump of RGB and ZAHeB phase, which suggests that the helium flash can be responsible for moderate production of Li. This increment provides an observational evidence for the Li production due to helium-flash-induced mixing events.  Furthermore, we found that Li-rich HeB stars locate anywhere during the HeB stars' steady state phase of core He-burning. We suggest that either some special physical mechanisms should be required in helium flash for vigorous Li production or they are formed via other scenarios that could result in random distributions in HeB phase. In addition, during the HeB stage, we found the depletion feature of Li is not obvious 
for stars with $A_{\rm Li}$ less than 1.5\,dex, which would be a visible manifestation of retreat of convective envelope as well as the inefficient deep-mixing process inside HeB stars.

\vspace{7mm} \noindent {\bf Acknowledgments}
We thank the referee for the thorough reviews which have helped us to improve the manuscript. This work is supported by National Key R\&D Program of China No.2019YFA0405502, the National Natural Science Foundation of China under grant Nos. 12090040, 12090044, 12022304,  U2031203, 11833006, 11973052,  11973042, 11988101, 11890694 and U1931102. H.-L.Y. acknowledges support from the Youth Innovation Promotion Association of the CAS (id. 2019060), and NAOC Nebula Talents Program. We acknowledge the entire \kepler{} team and everyone involved in the \kepler{} mission. Funding for the \kepler{} Mission is provided by NASA's Science Mission Directorate.
Guoshoujing Telescope (the Large Sky Area Multi-Object Fiber Spectroscopic Telescope, LAMOST) is a National Major Scientific Project built by the Chinese Academy of Sciences.
Funding for the project has been provided by the National Development and Reform Commission. LAMOST is operated and managed by the National Astronomical Observatories, Chinese Academy of Sciences.

\clearpage

\bibliography{ms2021-0830.bib}

\label{lastpage}

\end{document}